\documentclass[%
 aip,
 apl,
 amsmath,amssymb,
 reprint,%
]{revtex4-2}

\usepackage{graphicx}
\usepackage{dcolumn}
\usepackage{bm}
\usepackage[colorlinks=true,linkcolor=blue]{hyperref}

\usepackage[utf8]{inputenc}
\usepackage[T1]{fontenc}
\usepackage{mathptmx}
\usepackage{etoolbox}

\begin{document}

\preprint{AIP/123-QED}

\title{Magnetic metamaterials by ion-implantation}
\author{Christina Vantaraki}
\email{christina.vantaraki@physics.uu.se}
\affiliation{Department of Physics and Astronomy, Uppsala University, Box 516, 75120 Uppsala, Sweden}

\author{Petter Ström}
\affiliation{Department of Physics and Astronomy, Uppsala University, Box 516, 75120 Uppsala, Sweden}

\author{Tuan T. Tran}
\affiliation{Department of Physics and Astronomy, Uppsala University, Box 516, 75120 Uppsala, Sweden}

\author{Matías P. Grassi}
\affiliation{Department of Physics and Astronomy, Uppsala University, Box 516, 75120 Uppsala, Sweden}

\author{Giovanni Fevola}
\affiliation{Center for X-ray and Nano Science CXNS, Deutsches Elektronen-Synchrotron DESY, Notkestr. 85, 22607 Hamburg, Germany}

\author{Michael Foerster}
\affiliation{ALBA Synchrotron Light Facility, 08290-Cerdanyola del Valles, Barcelona, Spain}

\author{Jerzy T. Sadowski}
\affiliation{Center for Functional Nanomaterials, Brookhaven National Laboratory, Upton, NY 11973, USA}

\author{Daniel Primetzhofer}
\affiliation{Department of Physics and Astronomy, Uppsala University, Box 516, 75120 Uppsala, Sweden}

\author{Vassilios Kapaklis}
\email{vassilios.kapaklis@physics.uu.se}
\affiliation{Department of Physics and Astronomy, Uppsala University, Box 516, 75120 Uppsala, Sweden}

\begin{abstract}

We present a method for the additive fabrication of planar magnetic nanoarrays with minimal surface roughness. Synthesis is accomplished by combining electron-beam lithography, used to generate nanometric patterned masks, with ion implantation in thin films. By implanting $^{56}$Fe$^{+}$ ions, we are able to introduce magnetic functionality in a controlled manner into continuous Pd thin films, achieving 3D spatial resolution down to a few tens of nanometers. Our results demonstrate the application of this technique in fabricating square artificial spin ice lattices, which exhibit well-defined magnetization textures and interactions among the patterned magnetic elements.\end{abstract}

\maketitle

Magnetic nanoarrays  - {\it metamaterials} - have gained significant attention in recent years, both as systems of fundamental interest  \cite{HeydermanLJ2013Afsn, Gilbert:2016cn, nisoli2017deliberate, Rougemaille:2019ef, Skjarvo}, as promising candidates for applications in neuromorphic computing \cite{PappAdam2021Nnnu, Jensen.J, GartsideJackC2022Rtar}, as well as in magnonic and spintronic devices \cite{Gliga:2013jw, Iacocca:2016gb, Marrows_spintronics_2024}. The scalability and versatility of these systems have opened pathways for exploring emergent physics \cite{HeydermanLJ2013Afsn, nisoli2017deliberate} and expanding their potential applications. However, the topography of these magnetic nanoarrays, resulting from the nanopatterning process, can present significant challenges. A notable example is the observed strong structure-related photon scattering, which can obscure scattering resulting from magnetic ordering in the arrays \cite{woods_orbital_PRL2021}. Given that the photon scattering properties of these systems could play a crucial role in information technology applications \cite{Chen:2019eq}, or in x-ray metasurfaces for manipulating photon angular and orbital momentum \cite{woods_orbital_PRL2021}, it is essential to explore methods for developing truly planar arrays that maintain flexibility in material selection and precise spatial control. An additional motivation for investigating such processing stems from recent advances suggesting that magnetic metamaterials hold significant potential as physical substrates for computation \cite{Grollier_review_2020, GartsideJackC2022Rtar}. The possibility of integrating these materials with CMOS technology or magnetic random-access memory (MRAM) architectures highlights the need for expanding available fabrication methods \cite{Non_boolean_2016}.

To this end, ion-beam techniques offer a promising pathway. Previously, Focused Ion Beam (FIB) was used for creating magnetic arrays. FIB is a commonly used technique, where a beam of focused ions is directed towards a target and raster-scans over it to create a pattern.
FIB can be used with various processing techniques, such as FIB milling or implantation.
FIB milling uses ions to sputter material generating patterns of pillars. The topography of the patterns can be reduced by performing FIB implantation \cite{FassbenderJ.2008Mpbm}. 
However, the ion species available for FIB instruments are limited, while the serial nature of FIB makes it difficult to achieve large-scale patterns \cite{HoflichKatja2023Rffi}. 
Another approach to creating flat magnetic metamaterials involves ion-implantation through a physical mask. Since the pioneering work of \citet{ChappertC.1998PPMM}, magnetic patterns have been engineered by locally altering the magnetic properties of a film via ion-implantation \cite{EhresmannA2004Imps, Theis-BrohlKatharina2006Eiia, FassbenderJ.2008Mpbm, FassbenderJ2009Ials, MuscasG.2018Bntt}. 
However, it was not until recently that magnetic features were additively implanted \cite{StrmPetter2024PIoF}. Employing commercial masks, such as TEM grids, flat magnetic patterns were created, although the feasible designs remain constrained by the limitations of the available mask configurations.

\begin{figure}[h!]
\includegraphics[width=1\linewidth]{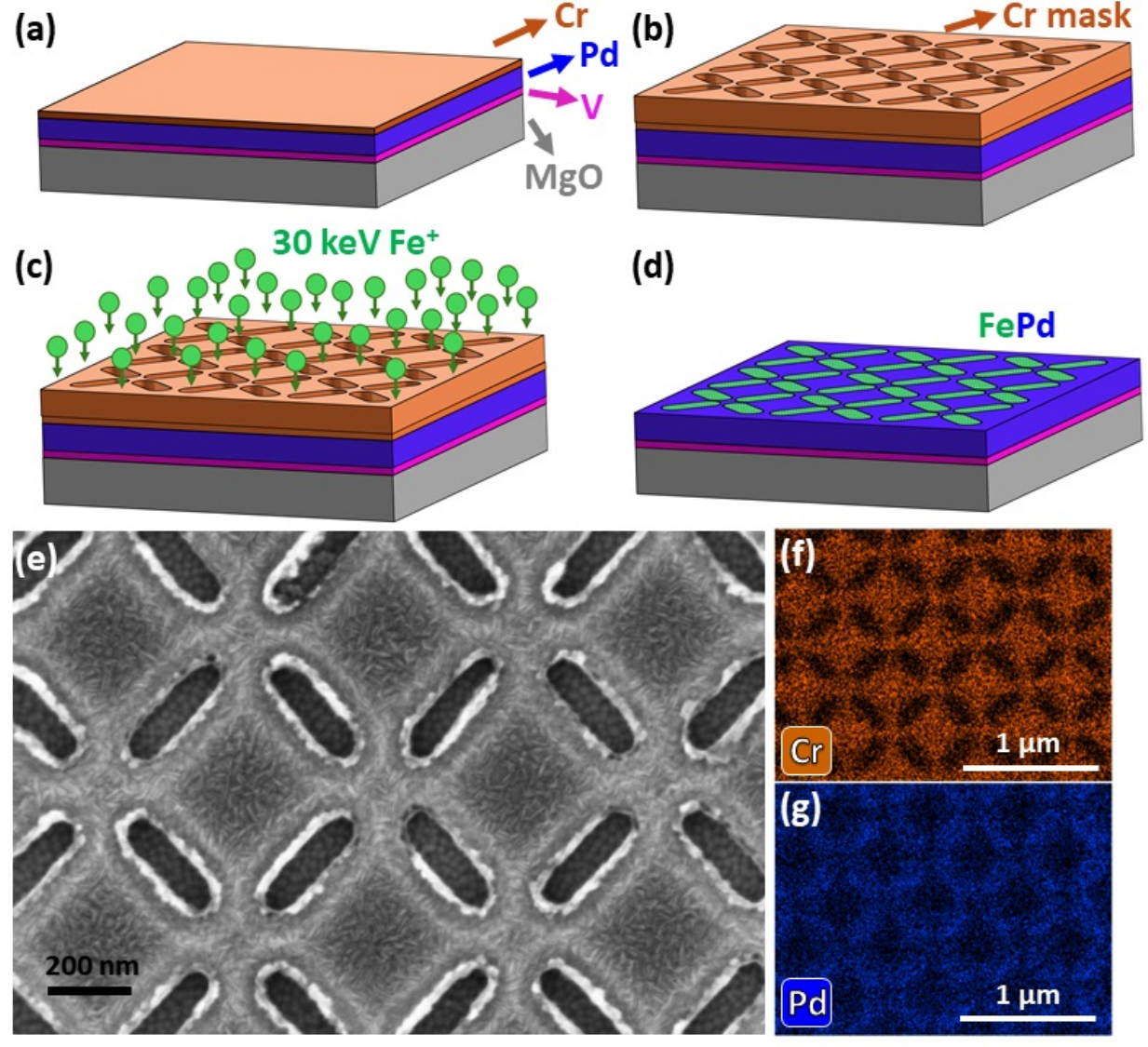}
\caption{(a-d) Fabrication process of implanted nanostructures: (a) Deposition of Pd film on a MgO substrate with V as adhension layer and Cr as capping layer, (b) Fabrication of the patterned Cr mask on top of the film, (c) Fe$^{+}$ implantation through the patterned mask, (d) Chemically removal of the Cr material. (e) SEM and (f,g) EDS images of the Pd film with the patterned Cr mask on top.}
\label{Figure1}
\end{figure}

In this work, we fabricate a patterned mask on top of a film through which magnetic patterns can be created using ion-implantation. We achieve this, combining the pattern design flexibility of modern nanolithography techniques with standard ion-implantation methods. 
Here the mask is the result of a serial-patterning in order to demonstrate the feasibility of the process. However, a parallel fabrication of the mask can be achieved by using extreme-ultraviolet lithography.
In the present work, $^{56}$Fe$^{+}$ ions are implanted into a Pd film through a Cr patterned mask, yielding flat ferromagnetic Fe$_x$Pd$_{100-x}$ (where $x$ stands for at.\%) nanostructures embedded into the Pd matrix.
The patterned mask enables the design of architectures with geometries and shapes that cannot be achieved with commercial masks such as TEM grids, while the dimensions of the building blocks can be reduced to the scale of a few tens of nanometers. 
The specific magnetic metamaterial architecture of choice is an Artificial Spin Ice (ASI) lattice due to the wide interest it has attracted and the potential to be implemented in applications \cite{Skjarvo, HeydermanLJ2013Afsn,nisoli2017deliberate,HeydermanLauraJ.2021MmsF, Chen:2019eq, woods_orbital_PRL2021, GartsideJackC2022Rtar}.

A 600~Å Pd film with a 50~Å V film adhesion layer was deposited on a MgO substrate using DC magnetron sputtering. Subsequently, 60~Å Cr were deposited on top of the Pd film by thermal evaporation, to act as a protection layer during the Fe$^{+}$ implantation (Fig. \ref{Figure1} (a)). 
A patterned Cr mask was prepared on top of the film by means of electron-beam lithography (EBL). More details are presented in the Supplementary Material. 
The patterned mask consists of stadium-shaped holes with length of 290 nm, width of 80 nm and depth of 60 nm placed in a square ASI lattice with periodicity 480 nm (Fig. \ref{Figure1} (b)). 
Scanning Electron Microscope (SEM) and Energy Dispersive Spectroscopy (EDS) images of the film with the patterned mask on top are shown in Fig.\ref{Figure1} (e) and (f,g), confirming a well-formed mask structure.

The pattern defined in the Cr mask was transferred to the Pd film by $^{56}$Fe$^{+}$ implantation, with an energy of 30 keV and nominal fluence of 4$\times$10$^{16}$ ions/cm$^{2}$ (Fig. \ref{Figure1} (c)). The selected fluence is low enough to ensure that the sample is not heated significantly, thereby hindering diffusion of the implanted Fe in Pd film. This step was followed by the chemical removal of the Cr material (Fig. \ref{Figure1} (d)). Details can be found in the Supplementary Material. Fig. \ref{Figure2} (a) shows a SEM image of the implanted pattern, where Fe$_x$Pd$_{100-x}$ stadium-shaped structures of length 300 nm and width 100 nm have been created in the Pd film. The final dimensions of the implanted nanostructures are slightly larger than the corresponding holes of the mask due to lateral straggling as Fe$^{+}$ ions penetrate the Pd film.
The surface topography of the sample was mapped performing Atomic Force Microscopy (AFM) in contact mode. The AFM image in Fig. \ref{Figure2} (b) shows that the implanted nanostructures have low surface roughness. Factors that contribute to the roughness are the sputtered Cr atoms during implantation and the implanted Fe atoms in the Pd film. By extracting height profiles across a nanostructure from the AFM images (Fig. \ref{Figure2} (c)), we conclude that the roughness is on the order of 4--5 nm. Cross-sectional Transmission Electron Microscopy (TEM) EDS analysis was performed on the implanted patterns, as shown in Fig. \ref{Figure3} (a). 
The TEM image confirms the low surface roughness of the implanted regions. The EDS spectrum displays a signal for Fe with a width of 300 nm along the long-axis of the implanted elements, detected in the Pd film. This observation confirms Fe implantation according to the patterned mask design. The intensity of the Fe signal is strong close to the sample surface and decreases towards the substrate, indicating that the Fe has a concentration depth-profile, as expected \cite{StrmPetter2022Soft}. This concentration-depth profile confirms the limited diffusion of Fe into the Pd film. Furthermore, some Cr is detected in the Fe implanted areas, highlighting that some Cr atoms have been implanted from the capping layer due to knock-on collisions as a side effect of the implantation process. To determine the Fe depth-profile of the sample, Ion Beam Analysis (IBA) in the form of Time-of-Flight Elastic Recoil Detection Analysis (ToF -- ERDA)\cite{Encyclopedia}, was performed on a continuous film that was prepared and implanted next to the masked sample. This measurement confirmed that Fe has indeed a concentration profile in the Pd film, with its peak close to the sample surface and at 18.4\% atomic concentration for the selected fluence (Fig. \ref{Figure3} (b)). 

\begin{figure}[t]
\centering
\includegraphics[width=1\linewidth]{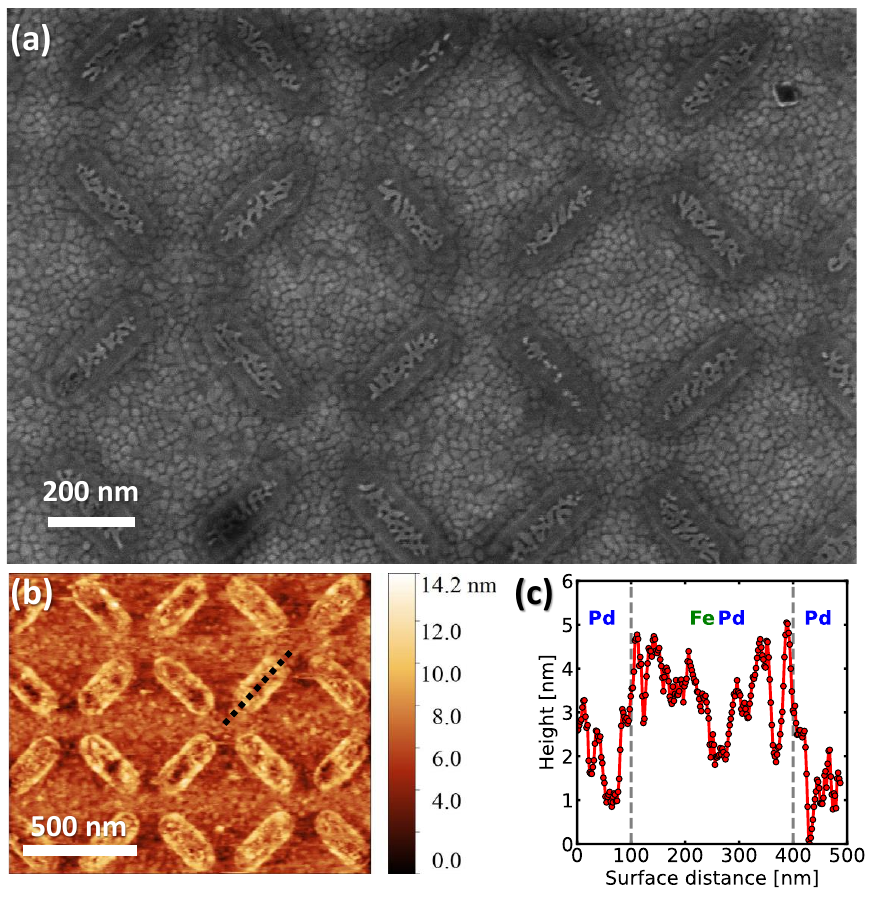}
\caption{(a) SEM and (b) AFM image of the implanted ASI lattice. (c) The height profile of a nanostructure extracted from the AFM image. The dashed line in (b) shows where the profile is extracted from.}
\label{Figure2}
\end{figure}

\begin{figure}[t]
\includegraphics[width=1\linewidth]{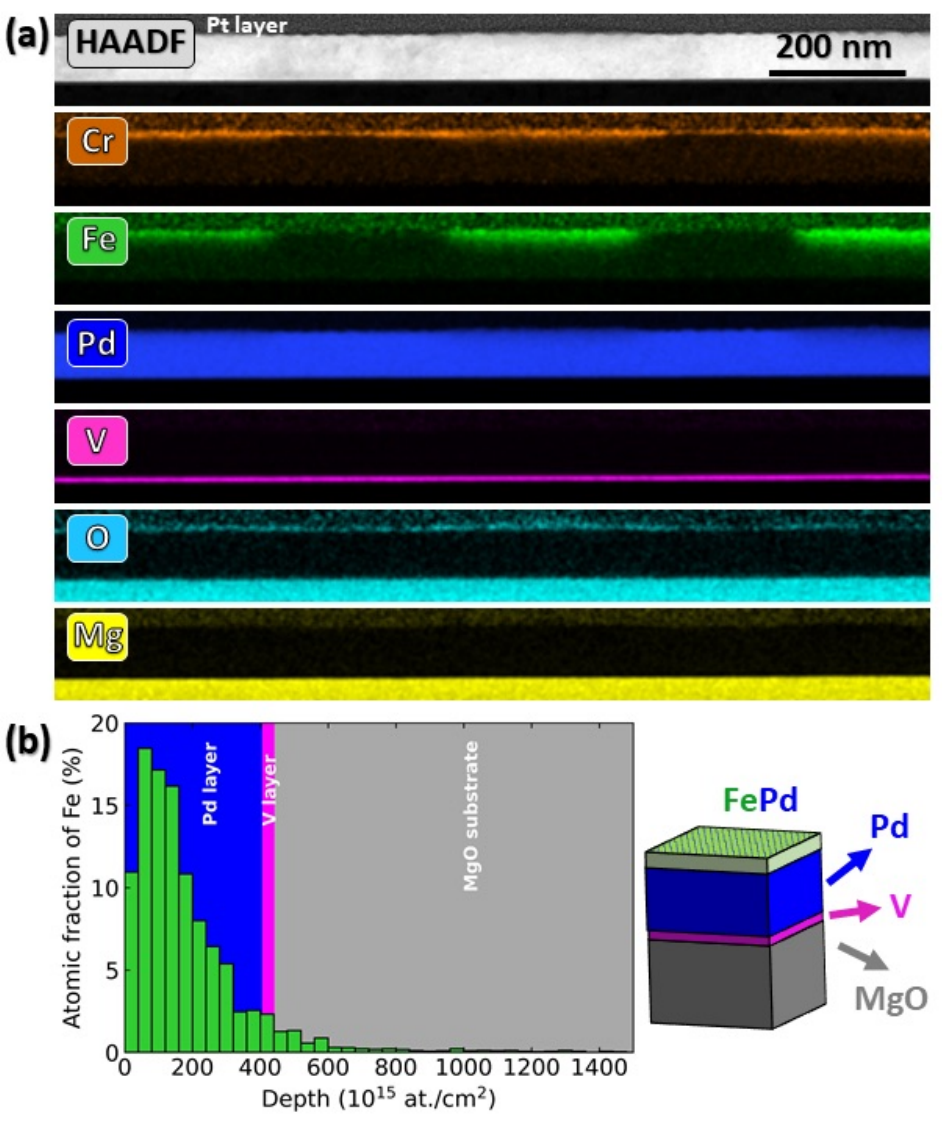}
\caption{(a) Cross-sectional TEM image and the corresponding EDS chemical mapping for Cr, Fe, Pd, V, O and Mg on the implanted pattern. A Pt layer has been deposited at the site of interest as protective layer for the TEM lamella preparation. All panels have the same magnification. The scale bar is shown in TEM image. (b) The Fe profile extracted from ToF-ERDA measurements on a continuous film that was prepared and implanted next to the patterned film. Due to the nature of this technique, the depth is measured in atoms/cm$^{2}$. Subsequently, the depth is converted to nanometers using the bulk density approximation. For Pd featuring bulk density, 10$^{15}$ atoms/cm$^{2}$ correspond to 1.47 Å. The extent of the Pd layer, V layer, and MgO substrate are added as background shading in the graph.}
\label{Figure3}
\end{figure}

\begin{figure}[t]
\includegraphics[width=1\linewidth]{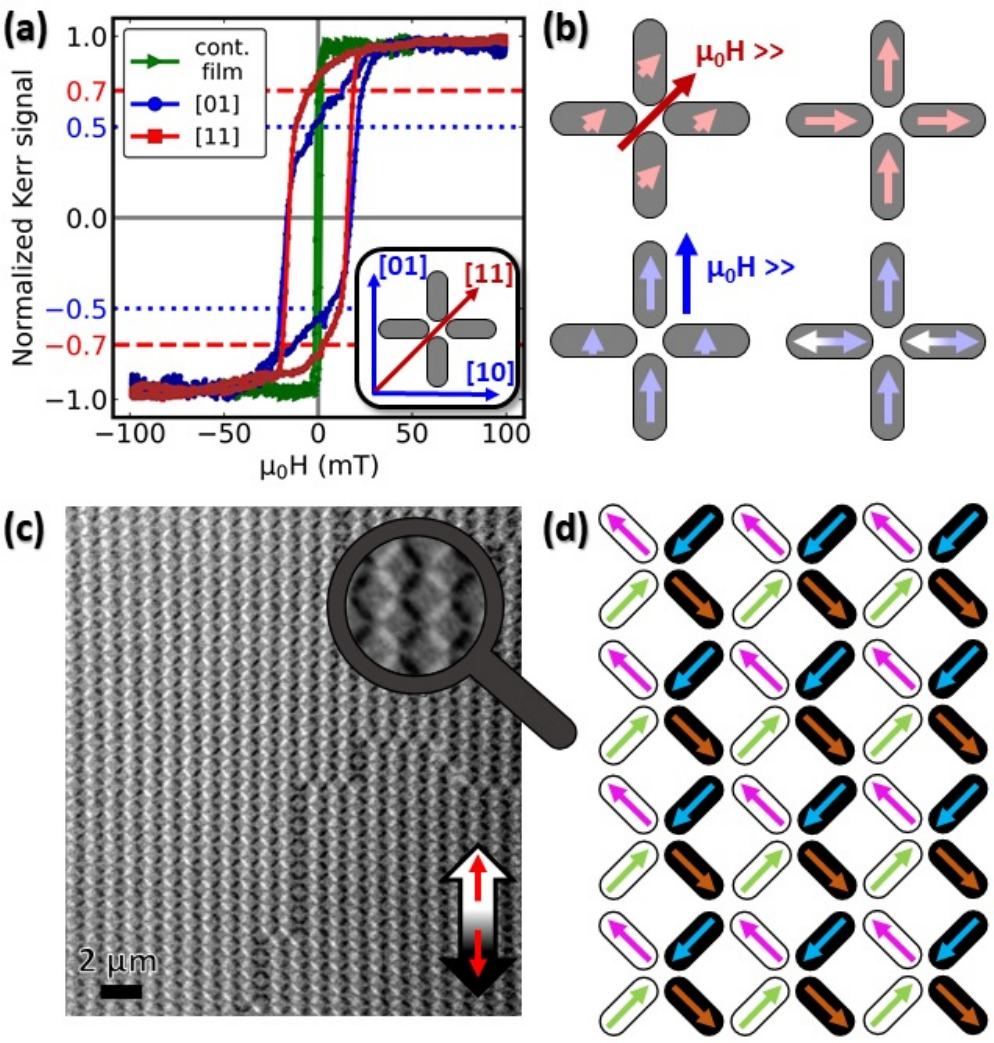}
\caption{(a) Magnetic hysteresis loop of the continuous (cont. film) and the patterned film ([01] and [11]), measured at room temperature by LMOKE setup. [01] and [11] indicate the direction of the applied magnetic field in respect to the lattice. The [01] and [11] directions are shown in the inset of the plot. (b) Saturation (left) and remanent (right) state of ASI lattice when the field is applied parallel to the [11] (top) and [01] (bottom) direction. (c) A representative PEEM-XMCD image of implanted ASI lattice. (d) A schematic of the ASI lattice showing Type I ordering. The arrows represent the direction of magnetization. The building blocks are white or black according to magnetic contrast obtained in a PEEM-XMCD experiment.}
\label{Figure4}
\end{figure}

The magnetic properties of the implanted films were studied using the Longitudinal Magneto-Optical Kerr Effect (LMOKE). Magnetic hysteresis loops were recorded at room temperature for an implanted continuous film and ASI array, and are shown in Fig. \ref{Figure4} (a). For the implanted ASI structure, the external magnetic field was applied in two directions, parallel to the [01] and parallel to the [11] directions (see inset Fig. \ref{Figure4} (a)).
The hysteresis loop of the continuous film shows a reduced remanence. This reduced remanence might be a consequence of Cr implantation and possible antiferromagnetic contribution due to that \cite{GRUNBERGP1986Lmse}. Similar effects have been reported for NiFe alloys implanted with Cr$^{+}$ \cite{FolksL2003Lmmo, FassbenderJ.2006Samm}.
For the implanted ASI structure, the remanent magnetization is altered depending on the direction of the field applied to saturate the array. When the array reaches the saturation state, the magnetic moments are aligned in the field direction (see Fig. \ref{Figure4} (b) left). After the removal of the field (remanent state), the shape of the elements forces the magnetization to be parallel with the long-axis of the elements, confining it in only two possible orientations. This restriction leads to a fully ordered remanent state after applying a magnetic field along the [11] direction (Fig. \ref{Figure4} (b) top right), yielding ideally a normalized remanent magnetization, with respect to the saturation magnetization, equal to $\sqrt{2}/2\approx0.707$. Similarly, a return to the remanent state after the application of a field along the [01] direction, leads to a full order for the elements parallel to [01] direction and a random order for the elements parallel to [10] direction (Fig. \ref{Figure4} (b) bottom right), yielding a normalized remanent magnetization equal to 0.5. This hysteretic behavior of the patterned array highlights the single magnetic domain structure of the implanted nanoelements. Conventional lithographic ASI lattices exhibited comparable behavior \cite{KapaklisVassilios2012Masi, SkovdalBjornErik2023Tewa}.

To directly illustrate the magnetic state of the implanted nanostructures, photoemission electron microscopy, employing x-ray magnetic circular dichroism (PEEM-XMCD) experiments were performed at the ESM (21-ID-2) beamline at the NSLS-II synchrotron and at the CIRCE (bl24) beamline at the ALBA synchrotron \cite{AballeLucia2015TAsL}. The PEEM-XMCD imaging was performed at room temperature in the absence of any external magnetic field, with a photon energy tuned to the Fe~L$_{3}$-edge (707 eV). A representative PEEM-XMCD image of an implanted ASI lattice is shown in Fig. \ref{Figure4} (c). The nanostructures are clearly displayed with white or black color, confirming their single magnetic domain state. 
Furthermore, the observed magnetic contrast confirms that the ASI lattice has obtained antiferromagnetic order (Type I), being the ground state of the square ASI lattice \cite{SchifferP2006A'ii}. This order establishes the presence of sufficiently strong magnetic interactions between the implanted nanostructures for the formation of long-range ordered domains. The formation of these extended domains, hints also towards an effective thermalization and relaxation process taking place during implantation, similar to temperature protocols which have been reported previously \cite{Morgan:2010di, KapaklisVassilios2014Tfia}.

The magnetic metamaterial fabrication process detailed here results in flat nanoarrays, which can be especially beneficial for magnetically controlled flat optical devices. \cite{woods_orbital_PRL2021, Steering_YIG}. 
This approach has the potential to enhance the steering of light beams by means of diffraction angles and energy range, offering a significant improvement over methods that rely on magnetic domain structures in yttrium-iron garnets \cite{Steering_YIG} or rare-earth transition metal alloys \cite{NanoLett_fs_CoGd}.
An additional advantage of this technique is the flexibility it provides in selecting the combination of implanted ion species and target materials. This flexibility enables the creation of material combinations that are not favoured in terms of thermodynamical stability, thereby expanding the range of possible materials for use in these applications. 
Moreover, the target material plays a crucial role in determining the overall magnetic properties of the metamaterials.
In this work, palladium was chosen for its high magnetic susceptibility \cite{Brodsky:1980bc, Hase:2014eg}. As a consequence, palladium obtains a strong induced magnetic moment from iron and exhibits ferromagnetic properties in a region that can extend several nanometers from the Fe interface \cite{Hase:2014eg}. This induced polarization effectively "smoothens'' the edges of the magnetic implanted nanostructures. Finally, this magnetic polarization effect enables interactions between the magnetic elements that can be tailored in ways that go beyond the capabilities of previous approaches, which were limited by nanolithography and geometric constraints\cite{Perrin:2016hj, Ostman:2018cp, Farhan:2019kd}.
Another possible area that can benefit from this technique is magnonic circuits. A flat ferromagnetic architecture in a non-ferromagnetic matrix can be an ideal candidate for fabricating magnonic busses. The lack of height step between the ferromagnetic and non-ferromagnetic parts could facilitate the patterning of the antennas needed in the magnonic circuits \cite{ChumakAV2017Mcfd}.

See the Supplementary Material for a detailed description of the fabrication process of the Cr patterned mask. We further present details on ToF – ERDA measurements and the MOKE system.

The authors would like to thank Johan Oscarsson and Mauricio Sortica at the Uppsala Tandem Laboratory for help with ion implantations. The authors are thankful for an infrastructure grant by VR-RFI (grant number 2019-00191) supporting the accelerator operation. The authors also acknowledge support from the Swedish Research Council (project no. 2019-03581). CV would like to thank David Muradas, Dr. Daria Belotcerkovtceva and Dr. Gopal Datt for fruitful discussions. CV further gratefully acknowledges ﬁnancial support from the Colonias-Jansson Foundation, Thelin-Gertrud Foundation, Liljewalch and Sederholm Foundation. We acknowledge Myfab Uppsala for providing facilities and experimental support. Myfab is funded by the Swedish Research Council (2020-00207) as a national research infrastructure. 
We would also like to thank Associate Professor Venkata Kamalakar Mutta for providing us chemicals for the process.
The PEEM-XMCD experiments were performed at CIRCE (bl24) beamline at ALBA Synchrotron with the collaboration of ALBA staff. M.F. acknowledges support from MICIN through grant number PID2021-122980OB-C54.
This research used resources of the National Synchrotron Light Source II and the Center for Functional Nanomaterials, U.S. Department of Energy (DOE) Office of Science User Facilities operated for the DOE Office of Science by Brookhaven National Laboratory under Contract No. DE-SC0012704. 
The authors are deeply grateful for the support from ReMade@ARI, funded by the European Union as part of the Horizon Europe call HORIZON-INFRA-2021-SERV-01 under grant agreement number 101058414 and co-funded by UK Research and Innovation (UKRI) under the UK government’s Horizon Europe funding guarantee (grant number 10039728) and by the Swiss State Secretariat for Education, Research and Innovation (SERI) under contract number 22.00187.

\hfill \break
\noindent
{\bf{DATA AVAILABILITY}}

The data that support the findings of this study are available from the corresponding authors upon reasonable request.

%

\pagebreak
\onecolumngrid
\newpage
\begin{center}
\textbf{\large Supplemental Material: Magnetic metamaterials by ion-implantation}
\end{center}
\setcounter{equation}{0}
\setcounter{figure}{0}
\setcounter{table}{0}
\setcounter{page}{1}
\makeatletter
\renewcommand{\theequation}{S\arabic{equation}}
\renewcommand{\figurename}{Supplementary FIG.}
\renewcommand{\thefigure}{{\bf \arabic{figure}}}
\renewcommand{\bibnumfmt}[1]{[S#1]}
\renewcommand{\citenumfont}[1]{S#1}
\renewcommand{\thepage}{S-\arabic{page}}

\section{Sample preparation} 
The film samples were prepared on a 10$\times$10$\times$1 mm$^{3}$ MgO substrate in an ultrahigh-vacuum (base pressure below 10$^{-7}$ Pa) DC magnetron sputtering system operating with Argon gas. A 600Å Pd film was deposited on top of 50Å V adhesion layer. To protect the Pd film from sputtering during the Fe$^{+}$ implantation process, the film was capped by 60Å Cr. The capping layer was deposited by thermal evaporation. 
After growth, a pre-patterned Cr mask was prepared on top of the film ahead of patterning. For this, a negative resist ma-N 2403 layer covered the film using a spin-coating process. To remove any solvents and achieve a smooth and flat surface of the resist, the film with the layer of resist on top was baked at 90 $^{\circ}$C for 60 sec. Using standard electron beam lithography technique, a geometrical pattern was directly transferred into the resist layer by means of a focused electron beam. 
The exposed film was placed into the developer solution ma-D 525 for 35 sec to dissolve the unexposed areas of the negative resist, followed by rinsing in deionized water for 10 min to stop the action of the developer.
Subsequently, 600Å Cr was deposited onto the film by thermal evaporation. Finally, a conventional lift-off process was used to remove the remaining resist together with the deposited material on top, yielding the pre-patterned Cr mask on top of the film. 
The unmasked and masked samples were implanted alltogether with 30 keV $^{56}$Fe$^{+}$ ions at the ion implanter of the Uppsala Tandem Laboratory. The total ion fluence was 4$\times$10$^{16}$ Fe ions/cm$^{2}$.
After the implantation, the Cr material was etched using diluted acid solution. The samples were placed in the acidic liquid for 50 sec, and rinsed thoroughly afterward with deionized water. 
The patterned area of the ASI structure studied with MOKE setup is 3$\times$3 mm$^{2}$, while the patches studied with PEEM-XMCD technque is 100$\times$100 $\mu$m$^{2}$.

\section{Time-of-flight elastic recoil detection analysis}
Time-of-flight elastic recoil detection analysis (ToF-ERDA) was carried out with a 36 MeV beam of $^{127}$I$^{8+}$, employing the detector system described in \citet{StrmPetter2016Acsa}, under standard measurement geometry. The angles of the trajectories of incoming and outgoing particles were 22.5$\pm$1.0$^{\circ}$ with respect to the sample surface normal. Conversion of the raw ToF-ERDA data to composition depth profiles was carried out with Potku \cite{ArstilaK.2014PNa}.

\section{Longitudinal Magneto-Optical Kerr Effect setup}
Magnetic hysteresis loops were recorded using the Magneto-Optical Kerr Effect in longitudinal configuration. The setup used a $p$-polarized laser beam with a wavelength of 660 nm that had a Gaussian profile with a spot diameter of approximately 2 mm. The incident laser beam was modulated using a Faraday cell before illuminating the sample at an angle. The sample was mounted on a holder between a quadrupole magnet. The reflected beam was passed through an analyzer (extinction ratio of 1:10$^{5}$), and the resulting intensity measured with a biased Si detector. The detector signal was fed to a pre-amplifier and a lock-in amplifier. The Kerr rotation was measured after applying a sinusoidal external magnetic field of frequency 0.1 Hz, and amplitude of 50 mT for the continuous films and 100 mT for the ASI lattices.

\end{document}